\documentclass[epj]{svjour}

\usepackage{graphicx}
\usepackage{fancyhdr}
\def\makeheadbox{{
 }}
\def\mytitle{My title} 
\def\myauthors{My name}  
\def\mytype{My type of session}
\def\mysession{My session}

\def\mytitle{MSSM precision physics at the {\it Z} resonance} 
\def\myauthors{Arne~M.~Weber}    
\def\mytype{Contributed Talk}    
\def\mysession{Colliders - SUSY Phenomenology}

\newcommand{\sweff}{\sin^2\theta_{\rm eff}}
\newcommand{\lsim}
{\;\raisebox{-.3em}{$\stackrel{\displaystyle <}{\sim}$}\;}

\usepackage{axodraw}

\begin{document}
\title{MSSM precision physics at the \boldmath{$Z$} resonance}
\author{S.~Heinemeyer\inst{1}
\and
W.~Hollik\inst{2}
\and
\underline{A.M.~Weber}\inst{2}
\thanks{\emph{Email:} Arne.Weber@mppmu.mpg.de}%
\and
G.~Weiglein\inst{3}
}                     
%
%
\institute{Instituto de Fisica de Cantabria (CSIC-UC), Santander,  Spain
\and Max-Planck-Institut f\"ur Physik (Werner-Heisenberg-Institut), F\"ohringer Ring 6, D--80805 Munich, Germany 
\and IPPP, University of Durham, Durham DH1 3LE, U.K.}
%
\date{}
\abstract{
LEP and SLC provide accurate data on the process $e^+e^-\to f\bar f$ at the $Z$~resonance.
The GigaZ option at a future linear $e^+e^-$ collider (ILC) will further improve these measurements.
As a consequence, theory predictions with sufficiently smaller errors are necessary in order to fully exploit the experimental accuracies and to derive indirect bounds on the scales of new physics.
Here we review the currently most accurate predictions of the $Z$~pole observables (e.g. $\sweff$, $\Gamma_Z$, $\sigma_{\rm had}^0$)
in the context of the Minimal Supersymmetric Standard Model (MSSM). These predictions contain the complete one-loop results including the full complex phase dependence, all available MSSM two-loop corrections, as well as all relevant Standard Model contributions.
} 
\maketitle
\section{Introduction}
\label{sec:intro}
$Z$~boson physics is well established as a cornerstone of the Standard Model
(SM)~\cite{LEPEWWG}.
Many (pseudo-) observables~\cite{Bardin:1997xq} 
have been measured with high accuracy at LEP and SLC using
the
processes (mediated at lowest order by photon and $Z$~boson exchange)
\begin{equation}
e^+e^-\to
f \bar f,\ \ \ \ \  f\neq e,
\label{epemtofermferm}
\end{equation}
at a center of mass energy $\sqrt{s} \approx M_Z$. 
In particular these are the effective leptonic weak mixing
angle at the $Z$~boson resonance, $\sweff$, $Z$~boson 
decay widths to SM fermions, 
$\Gamma(Z \to f \bar f)$, the invisible width, $\Gamma_{\rm inv}$,
the total width,
$\Gamma_Z$, forward-backward and left-right asymmetries, $A_{\rm FB}$ and
$A_{\rm LR}$, and the total hadronic cross section, $\sigma^0_{\rm had}$.
Here we focus on $\sweff$ and $\Gamma_Z$, as these two observables show the strongest sensitivity on effects due to virtual SUSY particles~\cite{ZObsMSSM}.

Together with the measurement of the mass of the $W$~boson, $M_W$, and
the mass of the top quark, $m_t$, the $Z$~pole 
observables have been instrumental in bounding the mass of the SM Higgs boson, 
the last free parameter of the model.
In a combined fit containing $Z$~pole observables, $W$~mass and $W$~decay width, 
the indirect constraints predict a SM Higgs boson mass of
$M_H = 76^{+33}_{-24}$~GeV, with an upper limit of 
$M_H\le 144$~GeV at the 95\%~C.L.~\cite{LEPEWWG}.

In order to fully exploit the high-precision measurements, the
theoretical uncertainty in the predictions of the (pseudo-) observables
should be sufficiently smaller than the experimental errors
(in view of the anticipated ILC precisions~\cite{moenig,gigaz}
for the $Z$~observables this is a particularly challenging task).
As a step in this direction we compute the currently most precise predictions for the observables at the $Z$~resonance~\cite{ZObsMSSM}. These contain the full one-loop result, all available higher order MSSM terms 
and all relevant SM contributions.
For the first time we include the full phase dependence at the one-loop level.

\section{The \boldmath{$Z$}~pole observables}
$Z$~pole pseudo observables are commonly defined and calculated in an effective coupling approach. This approach exploits the fact that the dominant contributions to the process $e^+e^-\to f\bar f$ at $s\sim M_Z^2$ stem from resonant $Z$~boson exchange diagrams. 
Non-resonant terms arise from photon exchange diagrams and box contributions, both of  which are accounted for as part of the unfolding procedure from the experimental data (see Refs.~\cite{Bardin:1997xq,ZObsMSSM} for details).
The electroweak radiative corrections can thus be absorbed into effective vector couplings, $g_V^f$, and axial vector couplings, $g_A^f$. These effective couplings are in turn used to define effective fermionic mixing angles (at Born level these coincide with the weak mixing angle, $\sin \theta_{\rm w}\equiv s_{\rm w}$)
\begin{equation}
\sweff^f := \frac{1}{4 |Q^f| }\Bigg(1 - {\rm Re}\Big[\frac{g_V^f}{g_A^f}\Big]\Bigg)=s_{\rm w}^2  {\rm Re}\left[\kappa_f\right]
\label{effmixangledef}
\end{equation}
 and partial decay widths
\begin{equation}
\Gamma_f = N_c^f \frac{\alpha}{3} M_Z \left(\left|g^f_V\right|^2   R_V^f +
  \left|g^f_A\right|^2  R_A^f\right).
\label{eq:Width}
\end{equation}
The latter are commonly expressed as~\cite{Bardin:1997xq}
\begin{equation}
\Gamma_f= N_c^f \bar\Gamma_0\left|\rho_f\right|\left(4(I_3^f-2Q^f s_{\rm w}^2\left|\kappa_f\right|)^2
  R_V^f +R_A^f\right), 
\label{eq:WidthalaHollik}
\end{equation}
with
\begin{equation}
\bar\Gamma_0=\frac{G_{\mu} M_Z^3}{24\sqrt{2}\pi}
\end{equation}
and $\rho_f$ derived from the axial vector couplings as detailed in Ref.~\cite{ZObsMSSM}.
In the above equations, $Q^f$, $I_3^f$ and $N_c^f$ stand for the charge, third isospin component and colour factor of the respective fermion~$f$. The radiation factors $R^f_{V,A}$ account for QED and QCD interaction of the final state fermions in the decay $Z\to f\bar f$. Tau lepton and more importantly bottom quark mass effects also enter via $R^f_{V,A}$.
The effective mixing angles defined in eq.~(\ref{effmixangledef}) are intimately realted to the forward backward and left right asymmetries measured at LEP and SLC. The total $Z$~boson width is obtained by summing over all partial widths~$\Gamma_f$
\begin{equation}
\Gamma_{Z} = \sum_{f=\nu,l,q} \Gamma_f \ .
\end{equation}

\section{One-loop result and higher order terms}
The computation of our MSSM predictions for the observables at the $Z$~resonance consists of two main steps: the computation of the full MSSM one-loop results, for the first time under consideration of {$\mathcal{CP}$}-violating complex MSSM parameters, and the inclusion of all available higher order contributions from SM and MSSM. Details of the calculations, which are briefly summarised in the following, can again be found in Ref.~\cite{ZObsMSSM}.

To accomplish our first goal we renormalise the $Z f\bar f$~vertex in the relevant parameters and compute all contributing MSSM one-loop graphs and counterterms.
All relevant Feynman graphs are calculated making use of the packages
{\tt FeynArts}~\cite{feynarts} and {\tt FormCalc}~\cite{formcalc}.  
As regularisation scheme dimensional reduction~\cite{dred} is used,
which allows a mathematically consistent treatment of
UV-divergences in supersymmetric theories at the one-loop level.
The inclusion of loop corrected Higgs masses and couplings in the complex MSSM 
is another new feature of our calculation.
Our implementation is performed
in accordance with the program {\tt FeynHiggs}~\cite{feynhiggs,mhiggsAEC} and the discussion in Ref.~\cite{mhcMSSMlong}. We furthermore resum the
$\tan \beta$~enhanced bottom Yukawa couplings following an effective coupling approach~\cite{deltamb2}.
For the first time we perform a full one-loop calculation for decay of the $Z$~boson into neutralino pairs, $Z\to \tilde\chi_1^0\tilde\chi_1^0$, which contributes to the invisible width and consequently also the total width of the $Z$~boson, provided $m_{\tilde \chi^0_1}<M_Z/2$.

In Ref.~\cite{ZObsMSSM} we give an exact description of the higher order terms which are included in our predictions for the observables at the $Z$~resonance.
Our philosophy regarding the inclusion of higher order terms is strongly influenced by fact that
the theoretical evaluation of the $Z$~pole observables in the SM is significantly more advanced than in the MSSM (see Ref.~\cite{dkappaSMbos2L} for a recent discussion of the state-of-the-art results in the SM). 
In order to obtain the most accurate predictions within the MSSM it is
therefore desireable to take all known SM corrections into
account. This can be done by writing 
the MSSM prediction for a quantity $x=g^f_{V,A},\rho_f,\kappa_f,\dots$ as
\begin{eqnarray}
x^{\rm MSSM} = x^{\rm SM}\big|_{M_H^{\rm SM} = M_{h_1}} + \underbrace{x^{{\rm MSSM} - {\rm SM}}}_{\equiv x^{\rm SUSY}} \ ,
\label{eq:obsSMSUSY}
\end{eqnarray}
where $x^{\rm SM}$ is the prediction in the SM with the SM Higgs boson
mass set to the lightest MSSM Higgs boson mass, $M_{H_1}$,
and $x^{{\rm MSSM}-{\rm SM}} \equiv x^{\rm SUSY}$  denotes the difference
between the MSSM and the SM prediction.
In order to obtain $x^{\rm MSSM}$ according to
eq.~(\ref{eq:obsSMSUSY}) we
evaluate $x^{{\rm MSSM}-{\rm SM}}$ at the level of 
precision of the known MSSM corrections, while for $x^{\rm SM}$
we use the currently most advanced result in the SM including all known
higher-order corrections. As a consequence, $x^{\rm SM}$ 
takes into account higher-order contributions which are only
known for SM particles in the loop, but not for their superpartners
(e.g.\ two-loop electroweak corrections to $\Delta\kappa$ beyond the
leading Yukawa contributions). Besides including all known higher order SM contributions, we also account for all available generic SUSY two-loop terms~\cite{dr2lA,drMSSMal2B}, which enter as universal corrections via the $\rho$~parameter.

\section{Dependence on complex parameters}
As already mentioned, $\sweff$ and $\Gamma_Z$ are the two $Z$~observables which show the strongest sensitivity on effects of new physics. We therefore focus on these two observables in the following. Analyses including $\sigma^0_{\rm had}, R_l, R_b$~etc.\ can be found in Refs.~\cite{ZObsMSSM,mastercode}.
Detailed numerical studies showed~\cite{ZObsMSSM} that the
dependence on the sfermion mass parameters is much stronger than on the
chargino/higgsino parameters. We therefore only investigate the dependence on the
phases of $A_t$ and $A_b$.
As for $\Delta r/M_W$~\cite{MWweber}, we find that
the effective couplings
$g_{V,A}^{f,(\alpha)}$ only depend on the absolute values $|X_t|$, $|X_b|$
of the off-diagonal entries in the $\tilde t$~and $\tilde b$~mass matrices,
where $X_t = A_t - \mu/\tan \beta$, $X_b = A_b - \mu\,\tan \beta$. Thus, the phases of 
$\mu$, $A_t$ and $A_b$  only enter in the combinations 
$(\phi_{A_{t,b}} + \phi_{\mu})$, giving rise to modifications of the squark
masses and mixing angles. It furthermore follows that the impact of $\phi_{A_t}$
($\phi_{A_b}$) on the sfermion 
masses is stronger for low (high) $\tan \beta$.

\begin{figure}[htb!]
\centering
\includegraphics[width=.45\textwidth,height=0.28325\textwidth,angle=0]{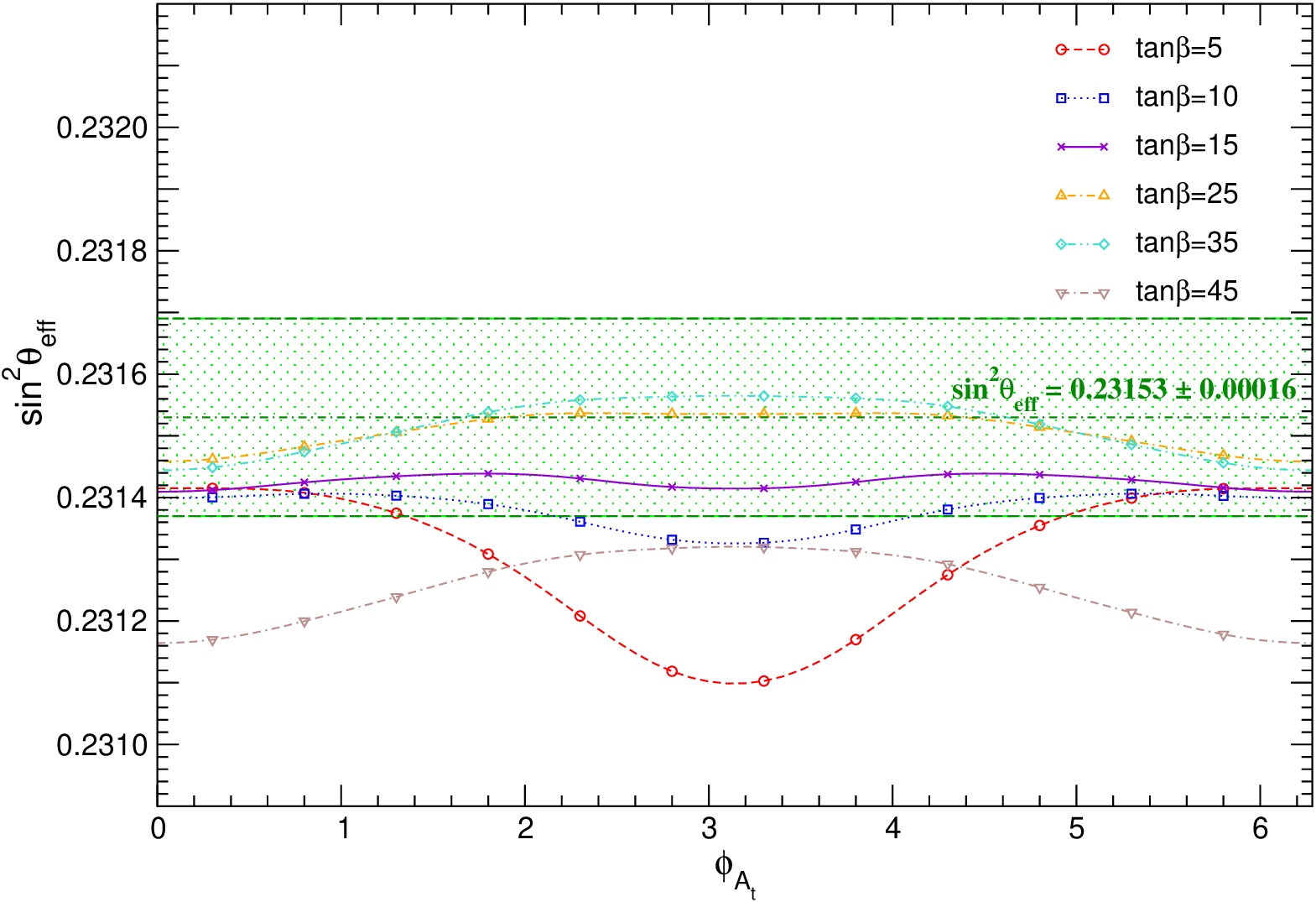}\\
\includegraphics[width=.45\textwidth,height=0.28325\textwidth,angle=0]{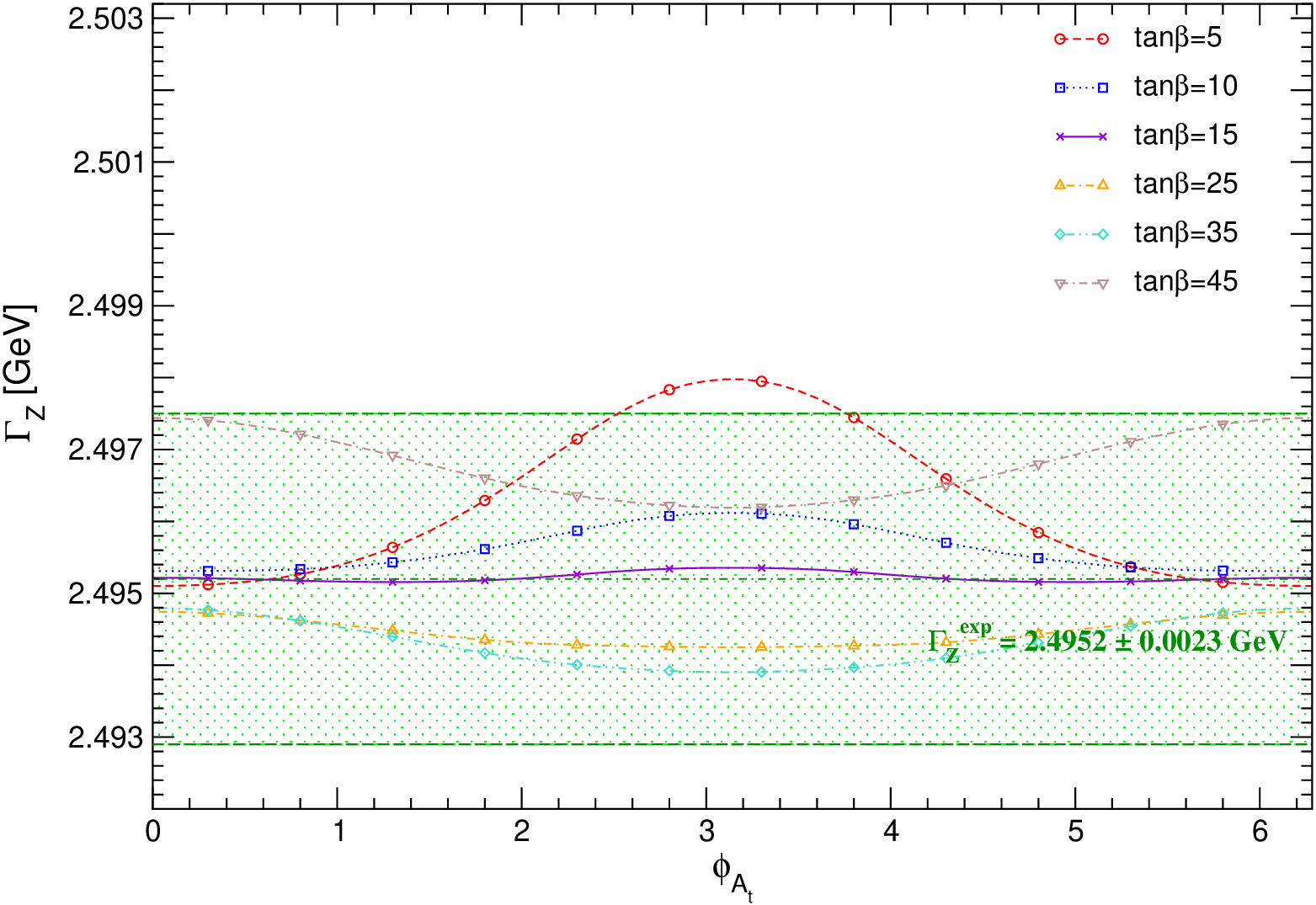}\\
\caption{Prediction for $\sweff$ and $\Gamma_Z$ as function of the
  phase of the trilinear coupling $A_t$. The other SUSY parameters are:
  $M_{\tilde f}= M_{H^\pm}= M_2 = m_{\tilde g}=500{\rm GeV}, A_\tau=A_t=A_b=\mu=1000~{\rm GeV},
  \phi_{\mu}=\phi_{M_1}=\phi_{M_2} =\phi_{\tilde g}=0,\phi_{A_b}=0$} 
\label{fig:phasesAt} 
\vspace*{-2em}
\end{figure}

In Figs.~\ref{fig:phasesAt}  we show $\sweff$ and $\Gamma_Z$ as a function of 
$\phi_{A_t}$ (with $\phi_{\mu} = \phi_{A_b} = 0$),
for different values of $\tan\beta$ (varied from $\tan\beta = 5$ to $\tan\beta = 45$).
Shown in the green-shaded bands are the current experimental values in 
  their $1\sigma$ range.
Using the same conventions as in Refs.~\cite{ZObsMSSM,MWweber}, the other parameters are set to  
$M_{\tilde f} = M_{H^\pm} = M_2 = m_{\tilde g} = 500$~GeV, 
$|A_{t,b,\tau}| = |\mu| = 1000$~GeV, $\phi_{M_1} = \phi_{M_2} =\phi_{\tilde g}= 0$.

As expected, the dependence of $\sweff$ and $\Gamma_Z$ on~$\phi_{A_t}$ is most pronounced for small $\tan\beta$. The variation of $\phi_{A_t}$ in this  case gives rise to a shift in the two precision observables by $1$--$2\sigma$. The effect becomes smaller for increasing $\tan\beta$, up to $\tan\beta=15$. 
On the other hand, for high $\tan\beta$ the lighter $\tilde b$~mass becomes rather small for the parameters chosen in Figs.~\ref{fig:phasesAt}, 
reaching values as low as about $100$~GeV for $\tan\beta=45$. This leads to a sizable shift of $\sim 1$--$2\sigma$ in the $Z$~observables already for vanishing phases. The slight rise in the dependence on $\phi_{A_t}$ for  $\tan\beta\geq25$ is due to the overall enlarged SUSY contributions which occur for  large $\tan\beta$ and the resulting low sbottom masses. We checked that the dependence on $\phi_{A_t}$ is in general significantly larger than the dependence  on $\phi_{A_b}$.

\section{MSSM parameter scans}
\label{subsec:scans}
We now investigate the behaviour 
of $\sweff$, the $Z$~observable most sensitive to higher order corrections, by scanning over a broad range of the SUSY parameter space.
The SUSY parameters given in Tab.~\ref{tab:scanrange} are varied independently of each other, within the given range, in a random parameter scan.
\begin{table}[tb!]
\begin{center}
\begin{tabular}{|c |c |c|}
\hline
{\bf  }& {\bf Parameter }& {\bf Range} \\
\hline
{ sleptons} & $M_{L_i}, M_{E_i}$ & $100$ to $2000~{\rm GeV}$  \\
\hline
{ squarks} & $M_{Q_i}, M_{U_i}, M_{D_i} $ &  $100$ to $2000~{\rm GeV}$  \\
              &  $A_{t}, A_b$ &  $-2000$ to $2000~{\rm GeV}$  \\
\hline
{ gauginos}& $M_{1}, M_{2}$ & $100$ to $2000~{\rm GeV}$  \\
& $M_3$ & $195$ to $1500~{\rm GeV}$  \\
\hline
{ Higgs} &  $M_A$ & $90$ to $1000~{\rm GeV}$  \\
 & $\tan\beta$ &  $1.1$ to $60$\\
\hline
&  $\mu$ &  $-2000$ to $2000~{\rm GeV}$ \\ 
\hline
\end{tabular}
\end{center}
\caption[MSSM parameter ranges in unconstrained random parameter scans.]{MSSM parameter ranges in the unconstrained random parameter scan shown in Figs.~\ref{fig:Scans2} and \ref{fig:Scans3}. Family indices~$i$ run from $1$ to $3$, where the respective parameters are varied independently. All phases are set to zero.}
\vspace*{-2.em}
\label{tab:scanrange}
\end{table}
Unlike Refs.~\cite{mastercode,AllanachFit,ehoww}, where our results for the electroweak precision observables were already employed in CMSSM multiparameter  analyses, 
the scans in this section are entirely unconstrained, apart from the fact that we require the Higgs mass bounds and direct SUSY particle exclusions limits from LEP searches to hold.

\begin{figure}[htb!]
\begin{center}
\includegraphics[width=.45\textwidth,height=0.4\textwidth,angle=0]{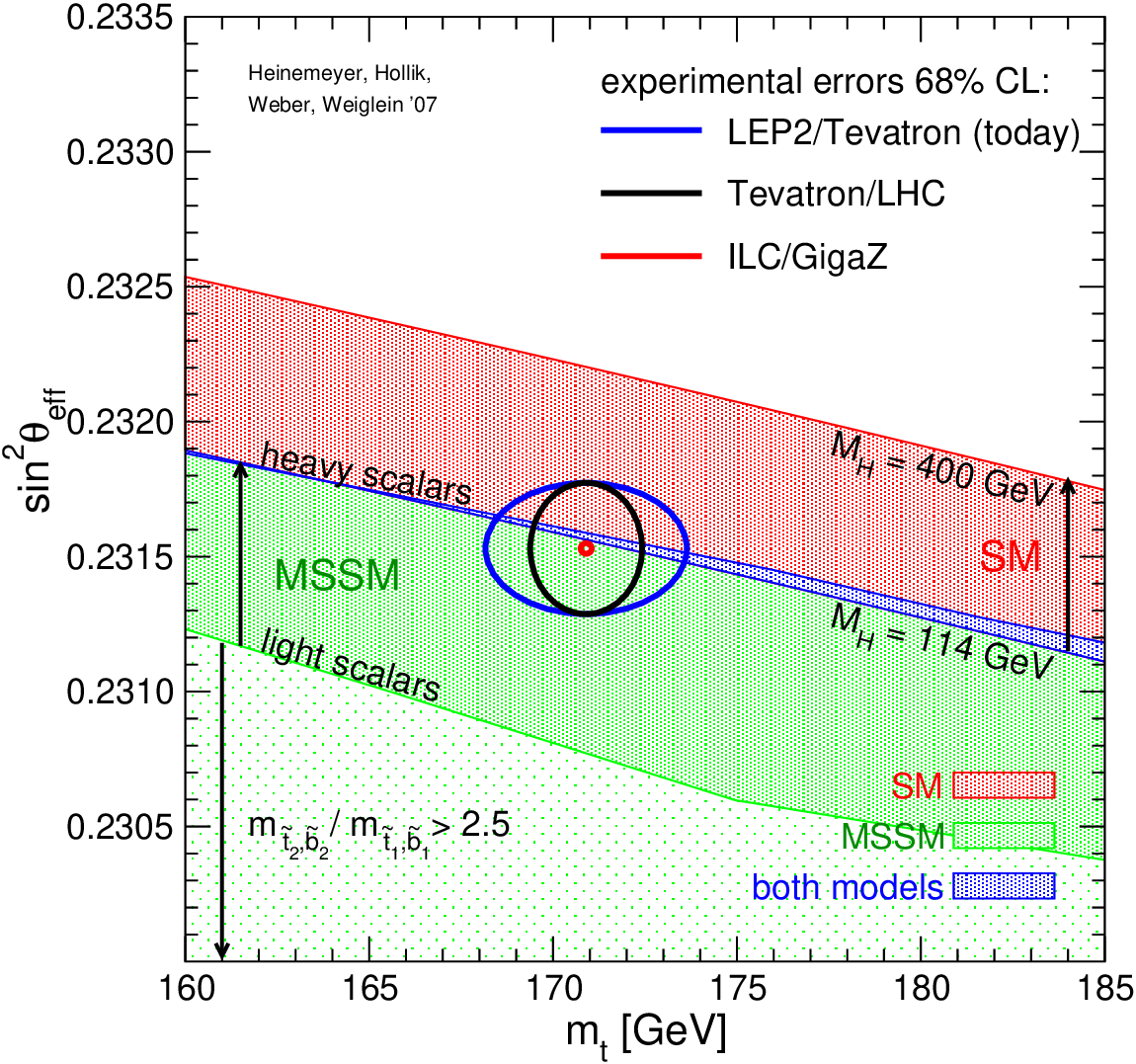}
\begin{picture}(0,0)
\CBox(-175,170)(-125,183){White}{White}
\end{picture}
\end{center}
\caption{Unconstrained MSSM random parameter scan for $\sweff$ as a function of $m_t$ over the
  ranges given in Tab.~\ref{tab:scanrange}. Todays 68\%~C.L.\ level ellipses 
  as well as future precisions, drawn around todays central value,  are
  indicated in the plot.}
\vspace*{-2em}
\label{fig:Scans2}
\end{figure}

The SM and the MSSM predictions for $\sweff$ as a function of $m_t$,
obtained from the scatter data with $m_t$ as an additional free parameter, are compared in Fig.~\ref{fig:Scans2}.
The predictions within the two models 
give rise to two bands in the $m_t$--$\sweff$ plane with only a relatively 
small overlap region (indicated by a dark-shaded (blue) area).
The allowed parameter region in the SM (the medium-shaded (red)
and dark-shaded (blue) bands) arises from varying the only free parameter 
of the model, the mass of the SM Higgs boson, from $M_{H}^{\rm SM} = 114$~GeV, the LEP 
exclusion bound~\cite{LEPHiggsSM}
(lower edge of the dark-shaded (blue) area), to $400$~GeV (upper edge of the
medium-shaded (red) area).
The very light-shaded (green), the light shaded (green) and the
dark-shaded (blue) areas indicate allowed regions for the unconstrained
MSSM. In the very light-shaded region at least one of the ratios
$m_{\tilde t_2}/m_{\tilde t_1}$ or $m_{\tilde b_2}/m_{\tilde b_1}$ exceeds~2.5
(we work in the convention that $m_{\tilde f_1} \le m_{\tilde f_1}$),
while the decoupling limit with SUSY masses of $\mathcal{O}({2 })$~TeV
yields the upper edge of the dark-shaded (blue) area. Thus, the overlap 
region between the predictions of the two models corresponds in the SM
to the region where the Higgs boson is light, i.e., in the MSSM allowed
region ($M_h \lsim 130 $~GeV~\cite{feynhiggs,mhiggsAEC}). In the MSSM it
corresponds to the case where all 
superpartners are heavy, i.e., the decoupling region of the MSSM.
The 68\%~C.L.\ experimental results
for $m_t=(170.9\pm1.8)$~GeV~\cite{mt1709} and $\sweff$ are indicated in the plot. As can be seen from Fig.~\ref{fig:Scans2}, the current 
experimental 68\%~C.L.\ region for 
$m_t$ and $\sweff$ is in good agreement with both models and does not 
indicate a preference for either of the two.
The prospective accuracies for the Tevatron/LHC 
($\delta\sweff^{\rm Tevatron/LHC} = 0.00016$, $\delta m_t^{\rm Tevatron/LHC} = 1 $~GeV)) and the future ILC with GigaZ option 
($\delta \sweff^{\rm ILC/GigaZ} = 0.000013$, 
$\delta m_t^{\rm ILC/GigaZ} = 0.1$~GeV) 
are also shown in the plot (using the current central values),
indicating the strong potential for a significant improvement of the
sensitivity of the electroweak precision tests~\cite{gigaz} (see Ref.~\cite{Erler:2007sc} for a recent review of the anticipated errors at future colliders).

\begin{figure}[htb!]
\begin{center}
\includegraphics[width=.45\textwidth,height=0.4\textwidth,angle=0]{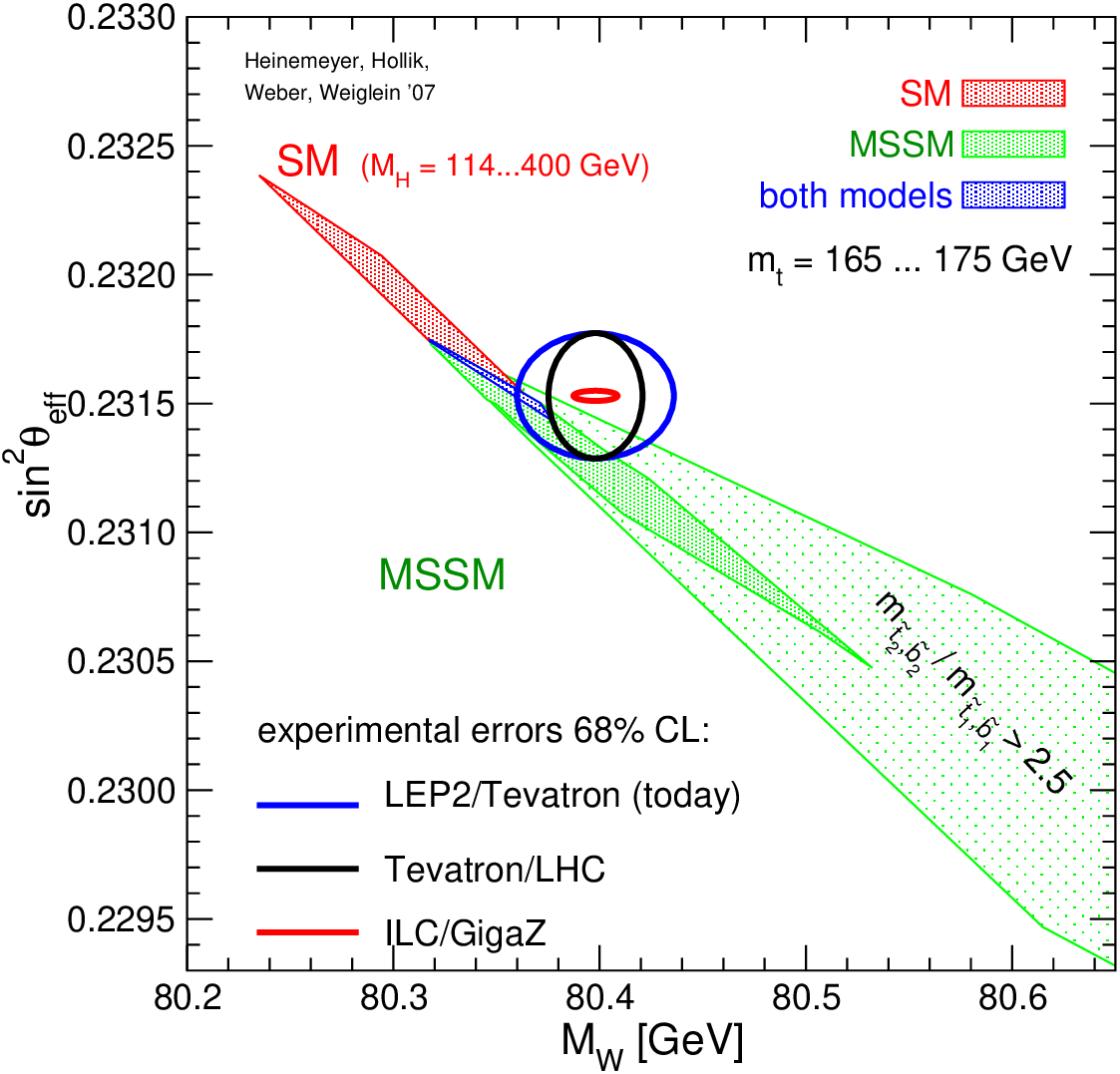}
\begin{picture}(0,0)
\CBox(-175,170)(-125,183){White}{White}
\end{picture}
\end{center}
\caption{Unconstrained MSSM random parameter scan over the ranges given in
  Tab.~\ref{tab:scanrange} and $m_t=165\dots175$~GeV. Shown is the combination of $M_W$ and
  $\sweff$. Todays 68\%~C.L.\ level ellipses 
  as well as future precisions, drawn around todays central value,  are
  indicated in the plot.}  
\vspace*{-2em}
\label{fig:Scans3} 
\end{figure}

In Fig.~\ref{fig:Scans3} we show the combination
of $M_W$~\cite{MWweber} and $\sweff$ with the top quark mass varied in the range of 
$165$~GeV to $175$~GeV.
The ranges of the other varied parameters and the colour coding 
are the same as in Fig.~\ref{fig:Scans2}.
The current 68\%~C.L.\ experimental results
for $M_W$ and $\sweff$ are indicated in the plot. The region of the SM
prediction
inside
todays 68\%~C.L.\ ellipse corresponds to relatively large $m_t$ values,
outside the current experimental range of 
$m_t=(170.9\pm1.8)$~GeV.
Thus, the combination of $M_W$ and $\sweff$ exhibits a slight preference
for the MSSM over the SM. 
The anticipated future
improvements in the measurements of $M_W$ and $\sweff$ are again indicated.

\end{document}